\documentstyle{aipproc}

\def\NPA{{\em Nucl. Phys.} A}

\def\PRL{\em Phys. Rev. Lett.}
\def\PRD{{\em Phys. Rev.} D}

\def\be{\begin{equation}}
\def\ee{\end{equation}}
\def\bea{\begin{eqnarray}}
\def\eea{\end{eqnarray}}
\def\bp{{\bf b}_\perp}
\def\qp{{\bf \Delta}_\perp}

\begin{document}
\title{Generalized Parton Distributions and the Dependence of Parton
Distributions on the Impact Parameter}

\author{Matthias Burkardt}
\address{Department of Physics\\
New Mexico State University\\
Las Cruces, NM 88003
}
\maketitle

\begin{abstract}
Generalized parton distributions (GPDs) provide a link between form
factors, parton distributions and other observables. I discuss the
connection between GPDs and parton distributions as a function of the
impact parameter. Since this connection involves GPDs in the limit of
vanishing skewedness parameter $\xi$, i.e. when the off-forwardness is
purely transverse, I also illustrate how to relate $\xi\neq 0$ data to
$\xi=0$ data, which is important for experimental measurements of these
observables. 
\end{abstract}

\section*{Introduction}

Deeply virtual Compton scattering experiments provide a useful
tool for probing 
off-forward or generalized parton distributions (GPDs) \cite{ji,ar}
\bea
\bar{p}^+\int \frac{dx^-}{2\pi} 
\langle p^\prime | \bar{q}(\frac{-x^-}{2}) \gamma^+
q(\frac{x^-}{2}) |p\rangle e^{ix\bar{p}^+x^-} 
&=&H^q(x,\xi,t)\bar{u}(p^\prime)\gamma^+ u(p)
\label{eq:off}\\
\nonumber
& &+E^q(x,\xi,t)\bar{u}(p^\prime)\frac{i\sigma^{+\nu}\Delta_\nu}{2M}u(p) 
, \\
\bar{p}^+\int \frac{dx^-}{2\pi} 
\langle p^\prime | \bar{q}(\frac{-x^-}{2}) \gamma^+\gamma_5
q(\frac{x^-}{2}) |p\rangle e^{ix\bar{p}^+x^-} 
&=&\tilde{H}^q(x,\xi,t)\bar{u}(p^\prime)\gamma^+\gamma_5 u(p)
\label{eq:offpol}\\
\nonumber
& &+\tilde{E}^q(x,\xi,t)\bar{u}(p^\prime)\frac{\gamma_5\Delta^+}{2M}u(p) 
, 
\eea
where $x^\pm = x^0\pm x^3$ and $p^+=p^0+p^3$ refer to the usual 
light-cone components, $\bar{p}=\frac{1}{2}\left(p+p^\prime\right)$,
$\Delta=p-p^\prime$, and $t\equiv \Delta^2$. The 
skewedness in Eqs. (\ref{eq:off},\ref{eq:offpol}) is defined as 
$\xi\equiv \frac{\Delta^+}{p^+}$. From the point of view of parton 
physics in the infinite momentum frame (IMF), GPDs have the
physical meaning of the amplitude for the process that a quark 
is taken out of the nucleon with longitudinal momentum fraction $x$ 
and then inserted back into the nucleon with a four momentum 
transfer $\Delta^\mu$ \cite{wally}.
GPDs play multiple roles and in a certain sense they interpolate
between form factors and
conventional parton distributions (PDs) \cite{ji,ar}:
for $\xi=t=0$ one recovers conventional PDs, i.e.
longitudinal
momentum distributions in the IMF, while
when one integrates $H^q(x,\xi,t)$ over $x$, one obtains a form factor,
i.e. the Fourier transform of a position space density 
(in the Breit frame!).
One of the new physics insights that one can learn from these
GPDs is the angular momentum distribution \cite{hood}.
Others include meson distribution amplitudes. \footnote{For a discussion
of this connection in the context of $QCD_{1+1}$ see Ref. \cite{mb:2d}.}

In this note, we will discuss the limit $\xi\rightarrow 0$, but $t\neq 0$, 
i.e. when the momentum transfer is purely transverse. In this limit,
the ``$E$-terms'' in Eqs. (\ref{eq:off}) and (\ref{eq:offpol}) drop out and
one finds
\bea
\int \frac{dx^-}{4\pi} 
\langle p^\prime | \bar{q}(\frac{-x^-}{2}) \gamma^+
q(\frac{x^-}{2}) |p\rangle e^{ixp^+x^-} 
&=&H^q(x,0,-{\bf \Delta}_\perp^2)
\label{eq:off2}
\\
\int \frac{dx^-}{4\pi} 
\langle p^\prime | \bar{q}(\frac{-x^-}{2}) \gamma^+\gamma_5
q(\frac{x^-}{2}) |p\rangle e^{ix\bar{p}^+x^-} 
&=&\tilde{H}^q(x,0,-{\bf \Delta}_\perp^2)
\label{eq:offpol2}, 
\eea
with ${p^+}^\prime = p^+$. Eqs. (\ref{eq:off2}) and (\ref{eq:offpol2})
very much resemble the definitions for ordinary twist-2 PDs,
with the only difference being the fact that the $\perp$ momenta of the
initial and final state are not the same. The situation here is very analogous
to the relation between the forward and off-forward matrix elements of
a current, i.e. between a charge and charge form factor. 
The main difference is of course that the operator
entering the `form factor' in Eq. (\ref{eq:offpol2}) is
not the current operator, but the operator that measures
longitudinal momentum distributions. From this analogy,
and since charge
form factors have the physical interpretation of the Fourier transform
of the position space charge distribution, it is natural to expect a similar
interpretation also for GPDs. 

\section*{GPDs for $\xi=0$}

In the following we will use a light-front (LF) Fock expansion to
represent GPDs for $\xi=0$ 
as overlap integrals between LF wave functions $\Psi_N(x,{\bf k}_\perp)$ 
summed $\left(\sum_N\right)$ over Fock components \cite{kroll1}
\footnote{Note that this is an exact expression provided
one knows the $\Psi_N$ for {\it all} Fock components}
\be
H^q(x,0,-{\bf \Delta}_\perp^2) = \sum_N \sum_j
\int \left[dx\right]_N \int \left[d^2{\bf k}_\perp\right]_N
\delta(x-x_j) \Psi^*_N(x,{\bf k}_\perp^\prime)
\Psi_N(x,{\bf k}_\perp),
\ee
where $\sum_j$ denotes the sum
over all quarks with flavor $q$ in that Fock component and
${\bf k}_{\perp, i}^\prime = {\bf k}_{\perp, i} - x_i {\bf \Delta}_\perp$ for
$i\neq j$, while 
${\bf k}_{\perp, j}^\prime = {\bf k}_{\perp, j} +(1-x_j) {\bf \Delta}_\perp$.
Upon switching to the coordinate representation in the $\perp$ direction
\be
\Psi_N(x,{\bf k}_\perp) = \int 
\frac{\left[d{\bf b}_\perp\right]}{(2\pi)^N}
e^{-i{\bf k}_\perp\cdot{\bf b}_\perp} \tilde{\Psi}_N(x,{\bf b}_\perp)
\ee
it is straightforward to see that
\be 
H^q(x,0,-{\bf \Delta}_\perp^2) = \int \left[dx\right]
\int \left[d{\bf b}_\perp\right]
\sum_N \sum_j \delta(x-x_j)
e^{i{\bf \Delta}_\perp\cdot({\bf b}_{\perp,j}-{\bf R}_\perp)}
\left|\tilde{\Psi}_N(x,{\bf b}_\perp)\right|^2.
\label{eq:density}
\ee
where we have introduced the {\it $\perp$ center of momentum}
\be
{\bf R}_\perp \equiv \sum_i x_i {\bf b}_{\perp,i} .
\label{eq:rperp}.
\ee
Eq. (\ref{eq:density}) illustrates that GPDs for $\xi=0$ can be interpreted
as Fourier transforms of {\it impact parameter dependent PDs}
\bea
H(x,0,-\qp^2)&=& \int d^2{\bf r}_\perp q(x,{\bf r}_\perp) 
e^{-i\qp\cdot{\bf r}_\perp}
\nonumber\\
\tilde{H}(x,0,-\qp^2)&=& \int d^2{\bf r}_\perp \Delta q(x,\bp) 
e^{-i\qp\cdot{\bf r}_\perp}
\label{eq:impact}
\eea
where for example (again we suppress spin indices for simplicity)
\be
q(x,{\bf r}_\perp)= \sum_N \sum_j \int \left[dx\right]
\int \left[d{\bf b}_\perp\right]
\delta(x-x_j) \delta\left({\bf r}_\perp - 
\left({\bf b}_{\perp,j} - {\bf R}_\perp\right)
\right) 
\left|\tilde{\Psi}_N(x,{\bf b}_\perp)\right|^2.
\ee
Thus, GPDs, in the limit of $\xi\rightarrow 0$, 
allow a simultaneous determination of
the longitudinal momentum fraction and transverse impact parameter
of partons in the target hadron in the IMF.

Eq. (\ref{eq:impact}) is not only a re-derivation of the main result from 
Ref. \cite{me} using LF Fock wave functions, but it
also clearly illustrates that the impact parameter in the impact parameter 
dependent PDs entering Eq. (\ref{eq:impact}) is measured w.r.t. 
${\bf R}_\perp$.
There is a striking similarity between this observation and the fact that
the Fourier transform of form factors in nonrelativistic (NR) systems yields 
charge distributions measured w.r.t. ${\vec R}_{CM} = \sum_i m_i {\vec r}_i/M$.
This should not come as a surprise, since there is a
residual Galilei invariance under the purely kinematic $\perp$ boosts
in the LF framework
\be
x_i \longrightarrow x_i^\prime = x_i \quad, \quad \quad \quad\quad
{\bf k}_{i\perp} \longrightarrow {\bf k}_{i\perp}^\prime
\equiv {\bf k}_{i\perp} + x_i \Delta {\bf P}_\perp,
\label{eq:imfboost}
\ee
which resembles very much NR boosts
\be
{\vec k}_i \longrightarrow {\vec k}_i^\prime ={\vec k}_i
+ m_i \Delta {\vec v} =
{\vec k}_i+\frac{m_i}{M} \Delta {\vec P} .
\label{eq:nrboost}
\ee
The above observation about the $\perp$ center of momentum
has one immediate consequence for the $x\rightarrow 1$
behavior of $q(x,\bp)$. Since the weight factors
in the definition of ${\vec R}_\perp$ are the momentum 
fractions, any parton $i$ that carries a large fraction $x_i$
of the target's momentum will necessarily have a $\perp$ 
position ${\vec r}_{i\perp}$ that is close to ${\vec R}_\perp$.
Therefore the transverse profile (i.e. the dependence on
$\bp$) of $q(x,\bp)$ will necessarily
become more narrow as $x\rightarrow 1$, i.e. we expect that
partons become very localized in $\perp$ position as 
$x\rightarrow 1$. By Fourier transform, this also implies
that the slope of $H(x,0,t)$ w.r.t. $t$ at $t=0$,
i.e.
\be
\langle {\vec b}^2_\perp \rangle \equiv
4 \frac{\frac{d}{dt}\left. H(x,0,t)\right|_{t=0} }
{ H(x,0,0) }
\ee
should in fact vanish for $x\rightarrow 1$!

\section*{Extrapolating to $\xi\rightarrow 0$}
From the experimental point of view, $\xi=0$ is not directly accessible
in DVCS since one needs some longitudinal momentum transfer in order to 
convert a virtual photon into a real photon.
There are several ways around this difficulty. 
First of all, one can access $\xi=0$ in real wide angle 
Compton scattering  \cite{ar3}. However, it should also be possible
to perform DVCS experiments at finite $\xi$ and to extrapolate
to $\xi=0$. 
This extrapolation is greatly facilitated by working with moments
since the $\xi$ dependence of the
moments of GPDs is in the form of polynomials \cite{wally}.
For example, for the even moments 
$H_{n}(\xi,t)\equiv \int_{-1}^1 \!\!\!dx x^{n-1} H(x,\xi,t)$
one finds\cite{hood}
\bea
H_{n}(\xi,t)
\label{eq:moment}
= A_{n,0}(t) + A_{n,2}(t) \xi^2+...+ A_{n,n-2}(t) \xi^{n-2}
+C_n(t)\xi^n, 
\eea
i.e. for example
\be
\int_{-1}^1 \!\!\!dx x H(x,\xi,t)
=A_{2,0}(t) + C_2(t)\xi^2.
\ee
Since the $H_n$ have a known functional dependence on $\xi$,
one can use measurements of the moments of GPDs at nonzero
values of $\xi$ to determine (fit) the ``form factors'' $A_{n,2i}(t)$ and 
$C(t)$. After determining these invariant form factors, one can 
evaluate Eq. (\ref{eq:moment}) for $\xi=0$, yielding
$
H_n(0,t) = A_{n,0}(t),
$
and the impact parameter dependence
of the $n-th$ moment of the parton distribution in the target reads
\be
q_n(\bp) \equiv \int_{-1}^1 \!\!dx x^{n-1} q(x,\bp)
= \int d^2q_\perp A_{n,0}(-\qp^2)e^{i\qp \bp}.
\ee
A very similar procedure can be applied to the moments of
spin dependent GPDs
\be
\tilde{H}_{n}(\xi,t)\equiv \int_{-1}^1 \!\!\!dx x^{n-1} 
\tilde{H}(x,\xi,t)
\label{eq:moment2}
= \tilde{A}_{n,0}(t) + \tilde{A}_{n,2}(t) \xi^2+...+ 
\tilde{A}_{n,n-1}(t) \xi^{n-1}. 
\ee
Similarly in the unpolarized case, one can extract
the $\frac{n+1}{2}$ form factors of 
the $n^{th}$ moment from measurements of $\tilde{H}$ for $\frac{n+1}{2}$
different values of $\xi$ (and the same values of $t$), yielding for the  
impact parameter dependence of the $n^{th}$ moment of the 
polarized PD
\be
\Delta q_n(\bp) \equiv \int_{-1}^1 \!\!dx x^{n-1} \Delta q(x,\bp)
= \int d^2{\bf \Delta}_\perp \tilde{A}_{n,0}(-\qp^2)e^{i\qp \bp}.
\ee
Of course, this procedure becomes rather involved for high moments,
but the steps outlined above seem to be a viable way of determining
the impact parameter dependence of low moments of parton distributions
from DVCS data.

\section*{Summary and outlook}
GPDs for $\xi\rightarrow 0$,
i.e. where the off-forwardness is only in the $\perp$
direction, can be identified with the Fourier transform of 
impact parameter dependent PDs. Here the impact parameter 
${\bf b}_\perp$.
is defined as the $\perp$ distance from the
center of (longitudinal) momentum in the IMF. This 
identification of GPDs with Fourier transforms of
impact parameter dependent PDs is very 
much analogous 
to the identification of the charge form factor with the
Fourier transform of a charge distribution in position space.

Although the $\xi\rightarrow 0$ limit of GPDs cannot 
be probed directly in DVCS, one can use the know polynomial 
$\xi$-dependence of the $x$-moments to extrapolate to $\xi=0$. 

Knowing the impact parameter dependence allows one to gain information
on the spatial distribution of partons inside hadrons and to obtain
new insights about the nonperturbative intrinsic structure of hadrons.
For example, the pion cloud of the nucleon is expected to contribute more
for large values of ${\bf b}_\perp$. 
Shadowing of small $x$ parton distributions,
is probably stronger at small values of ${\bf b}_\perp$ since
partons in the geometric center of the nucleon are more effectively
shielded by the surrounding partons than partons far away from the 
center. Geometric models for the small $x$ behavior of the PDs in the nucleon
suggest that polarized PDs may be more spread out in ${\bf b}_\perp$ than
unpolarized ones \cite{tt}.
These and many other models and intuitive pictures 
for the parton structure of hadrons give rise to predictions
for the impact parameter dependence of PDs that reflect the
underlying microscopic dynamics of these models. 
Our results may also play
an important role in the modeling of the $t$ dependence in GPDs, which
may in turn be relevant for fitting GPDs to experimental data for DVCS
amplitudes.
Finally, combining information about the impact parameter
dependence with information about longitudinal correlations 
in position space \cite{weise} 
may lead to further insights into non-perturbative hadron structure.

{\bf Acknowledgments:}
This work was supported by 
the DOE (DE-FG03-95ER40965) and by TJNAF. I would like to thank R.L. Jaffe,
S. Brodsky and R. Jakob for useful and encouraging comments.

\end{document}